\documentclass[pra,twocolumn,groupedaddress,showpacs,byrevtex]{revtex4}
\usepackage{graphics,amssymb,amsmath}

\newcommand{\sect}[1]{Sec.~\ref{#1}}
\newcommand{\fig}[1]{Fig.~\ref{#1}}
\newcommand{\eqn}[1]{Eq.~(\ref{#1})}
\newcommand{\eqns}[2]{Eqs.~(\ref{#1}) and (\ref{#2})}
\newcommand{\ket}[1]{| #1 \rangle}
\newcommand{\bra}[1]{\langle #1 |}
\newcommand{\braket}[2]{\langle #1 | #2 \rangle}
\newcommand{\floor}[1]{\lfloor #1 \rfloor}
\newcommand{\ceil}[1]{\lceil #1 \rceil}
\newcommand{\sinc}{\mathrm{sinc}}
\newcommand{\adag}{a^\dagger}
\newcommand{\be}{\begin{eqnarray}}
\newcommand{\ee}{\end{eqnarray}}

\begin{document}

\title{Generation of eigenstates using the phase-estimation algorithm}

\author{B.~C.~Travaglione}
 \email{btrav@physics.uq.edu.au}
 \homepage{www.physics.uq.edu.au/people/btrav/}
\author{G.~J.~Milburn}
 \affiliation{
  Centre for Quantum Computer Technology, University of Queensland,
  St. Lucia, Queensland, Australia
  }

\date{5 February 2001}
\begin{abstract}

The phase estimation algorithm is so named because it 
allows the estimation of the \emph{eigenvalues} associated with an operator. 
However it has been proposed that the algorithm can also be 
used to generate \emph{eigenstates}.  Here we extend this proposal for small 
quantum systems, identifying the conditions under which the phase 
estimation algorithm can successfully generate eigenstates. We then propose 
an implementation scheme based on an ion trap quantum computer. This 
scheme allows us to illustrate two simple examples, one in which the algorithm 
effectively generates eigenstates, and one in which it does not.

\end{abstract}

\maketitle

\section{Introduction}\label{intro}

Since the inception of quantum computation~\cite{Feynman82}, people in the
field have endeavored to find tasks which a quantum computer could perform
more efficiently than a classical 
computer~\cite{Deutsch92,Simon94,Shor94,Grover97a}. For a detailed
introduction into the field of quantum computation and information, see
\cite{Nielsen00}. The algorithm which has by far generated the most interest
is Shor's factoring algorithm~\cite{Shor94}, as it enables the cracking of the
RSA encryption system~\cite{rivest78}. Kitaev~\cite{Kitaev95} generalized
Shor's algorithm, showing how a quantum computer can generate an eigenvalue
of an arbitrary unitary operator (in the limit of a large number of qubits,
and not necessarily efficiently). Due to experimental difficulties, a large
scale quantum computer (if possible), will not be attainable for a number of
years. However, small-scale quantum computers are already 
available~\cite{Sackett00}. In this paper, we show how a version 
of the phase estimation algorithm can be implemented on a particular 
`small-scale' quantum computer, the ion trap quantum computer.

Given some unitary operator $U$ and an approximate eigenstate; the goal of the
phase estimation algorithm~\cite{Kitaev95,Cleve98} is to obtain an eigenvalue of
$U$ and leave the quantum system in the corresponding 
eigenstate~\cite{Abrams99b,Zalka96}.  
To accomplish this task, we shall need two quantum systems which can be coupled 
together. One, we shall call the index system, the other the target system. The 
index system is initially prepared in the state $\ket{0}$. After performing
the algorithm, the index system will store an eigenvalue of the target system 
operator, $U$.

Traditionally both the target and index systems have been qubit registers. In 
this paper the index system will remain a register of qubits, however we shall 
allow the target system to be an arbitrary $N$-dimensional quantum system, 
where $N$ may be equal to infinity. 
For a more generalized discussion of combining continuous and discrete quantum 
computation see \cite{Lloyd00}.

In \sect{phalgorithm} we briefly review the phase estimation algorithm then 
derive the analytical results which will allow us to characterize the 
algorithm's performance when using only a small number of qubits. 
In \sect{ionimp} we derive the Hamiltonians necessary 
to investigate the number and displacement operators in an ion trap, and 
contrast the algorithms effectiveness with respect to the two different 
operators.

\section{The Phase Estimation Algorithm}\label{phalgorithm}

In what follows, we shall assume that our index system is a register of $m$
qubits. First, we need to be able to perform the operation $\Lambda(U)$ on our
coupled system. $\Lambda(U)$ is completely described by defining its action on
the standard basis states of the index system, coupled to an arbitrary target
system state,
\be\label{lambdaU}
 \Lambda(U)\ket{j}_I\ket{\psi}_T
         &=& {\mathbb I}_I \otimes {U^j}_T \ket{j}_I\ket{\psi}_T \nonumber \\
     &=& \ket{j}U^{j} \ket{\psi} \quad \forall \ j \in \mathbb{Z}_M,
\ee
where $\mathbb{Z}_M = \{0,1,2,\dots,M-1\}$ and $M = 2^m$.
As in the last line of \eqn{lambdaU}, we shall continue to omit
the subscript notation when it is clear whether a ket or operator is
referring to the target or index system.
We begin the algorithm by initializing our quantum computer into the state
\be
 \ket{\Psi_0} = \ket{0}\ket{\psi}.
\ee
Performing a $\pi/2$ rotation of each qubit in the index
register results in the state
\be
 \ket{\Psi_1} = \frac{1}{\sqrt{M}} \sum^{M-1}_{j=0} \ket{j} \ket{\psi},
\ee
We now perform $\Lambda(U)$ on this state giving
\be\label{sumU}
  \ket{\Psi_2} = \Lambda(U)\ket{\Psi_1}
  &=& \frac{1}{\sqrt{M}} \sum^{M-1}_{j=0} \ket{j} U^j \ket{\psi}.
\ee

\begin{figure}[ht]
 \centering
 \scalebox{0.50}{\includegraphics{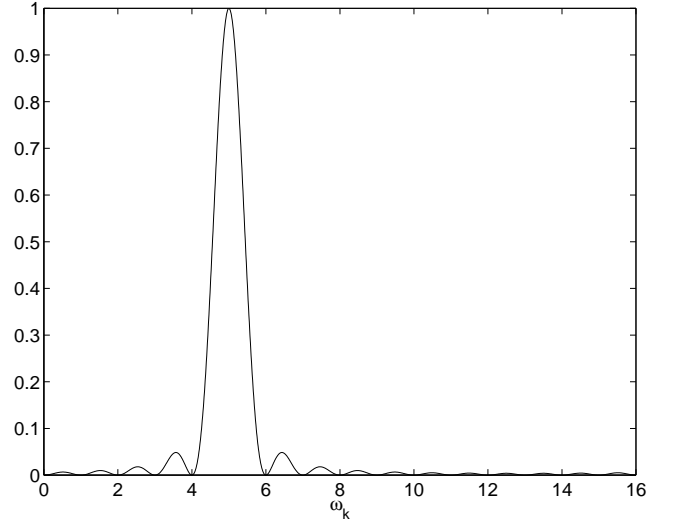}}
 \caption{A plot of $|f|$ as a function of $\omega_k$, with $M = 16$ and $j = 5$.}
 \label{fplot}
\end{figure}

The final steps 
in the algorithm are to perform the unitary quantum Fourier
transform~\cite{Coppersmith94} on the index register and measure this
register 
\footnote{By combining the QFT and measurement steps we
remove the need to perform any two qubit operations. The two qubit controlled
rotation operations can be replaced with measurement and conditional single 
qubit rotations. However for clarity we keep the description of these two steps
separate.}. 
However, before applying this transform we shall re-write
\eqn{sumU}.  First we replace $\ket{\psi}$ by it's representation as a sum of
eigenvectors of $U$,
\be\label{initialts}
 \ket{\psi} = \sum_k c_k \ket{\phi_k},
\ee
where $k$ sums over the dimensionality of the target system.
Hence the state $\ket{\Psi_2}$ can be written as
\be\label{rearr1}
  \ket{\Psi_2} &=&
  \frac{1}{\sqrt{M}} \sum^{M-1}_{j=0} \ket{j} U^j \sum_k c_k \ket{\phi_k}.
\ee
We shall write the eigenvalue associated with $\ket{\phi_k}$
as $e^{i\phi_k}$. Noting that $U^j$ applied to each eigenvector
$\ket{\phi_k}$ is simply $e^{ij\phi_k} \ket{\phi_k}$ and changing the order
of the summations we obtain
\be\label{rearr2}
  \ket{\Psi_2} = \sum_k c_k \frac{1}{\sqrt{M}} \sum^{M-1}_{j=0}
  \ket{j} e^{ij\phi_k} \ket{\phi_k}.
\ee
Lastly, for clarity we exchange the order of the systems, and
replace $\phi_k$ with $2\pi \omega_k/M$, where $\omega_k \in
\left[0,M\right)$,
\be\label{rearr3}
  \ket{\Psi_2} &=& \sum_k c_k \ket{\phi_k}
    \frac{1}{\sqrt{M}} \sum^{M-1}_{j=0} e^{2\pi i j \omega_k/M} \ket{j}.
\ee
It is now not hard to show that taking the quantum Fourier
transform of the index register results in the state
\be\label{finalstate}
  \ket{\Psi_3} &=& \sum_k c_k \ket{\phi_k}
     \sum^{M-1}_{j=0} f(\omega_k,j) \ket{j},
\ee
where
\be\label{phasefun}
 f(\omega_k,j) &=& \left\{
    \begin{array}{c@{\ :\ }l}
       \frac{1}{M} \frac{\sin(\pi \omega_k)}{\sin(\pi \frac{\omega_k - j}{M})}
           e^{\pi i (\omega_k - \frac{\omega_k - j}{M})} & \omega_k \not= j \\
       1 & \omega_k = j. \end{array}  \right.
\ee
As we will see shortly, it is helpful to note that
\be
 |f(\omega_k,j)| &\geq& |\sinc(\omega_k - j)|,
\ee
for all $\omega_k \in \left[0,M\right)$ and $j \in \mathbb{Z}_M$. 
A plot of $|f(\omega_k,j)|$ is shown in \fig{fplot} where $M = 16$ and $j$ has
been set to 5.

Finally, measuring the index register will, with high
probability, yield an approximate eigenvector. To understand this, let us begin
by looking at the most simplified case. Suppose for a moment, that
we have $\omega_k \in \mathbb{Z}_M$ for all $k$, then
\be
   f(\omega_k,j) &=& \delta_{\omega_k-j}.
\ee
Thus \eqn{finalstate} simplifies to
\be\label{simplestate}
  \sum_k c_k \ket{\phi_k}\ket{\omega_k}.
\ee
If we add the assumption that no two values of $k$
give the same $\omega_k$ (i.e. we have no degeneracy \footnote{Please note that
this unrealistic situation is not even \emph{possible} if our target system
has dimension greater than $2^m$.}) then upon measuring the index register,
we will obtain $\ket{\omega_k}$, and hence $e^{i\phi_k}$, with
probability $|c_k|^2$, and leave the target system in the
eigenstate $\ket{\phi_k}$.

Removing the assumption of zero degeneracy, measuring
the index register still allows us to obtain some eigenvalue
$e^{2\pi i j/M}$, however the target system is now left in the state
\be
 \frac{1}{\sqrt{\mathcal{N}}} \sum_{k'} c_{k'} \ket{\phi_{k'}}
\ee
where $k' = \{k : \omega_k = j \}$, and $\mathcal{N} = \sum_{k'} |c_{k'}|^2$ is
a normalization constant.

Finally, we shall remove the assumption that the $\omega_k$ must be elements
of $\mathbb{Z}_M$. The probability $P(j)$, of measuring the index register in
some basis state $\ket{j}$ is
\be\label{probj}
 P(j) &=& \sum_k |\bra{\phi_k} \braket{j}{\Psi_3} |^2 \nonumber \\
     &=& \sum_k |c_k f(\omega_k,j) |^2.
\ee
Having measured the index register to be in some state $\ket{j}$, the target
system is left in the state
\be\label{ampj}
 \ket{\psi'_j} &=& \sum_k c'_k \ket{\phi_k}, 
 	\quad c'_k = \frac{c_k f(\omega_k,j)}{\sqrt{\mathcal{N}}} 
\ee
where $\mathcal{N} = \sum_{k} |c_{k}f(\omega_k,j)|^2$.

In order to gain some useful information from
\eqns{probj}{ampj}, let us assume that our initial target
system state $\ket{\psi}$ is an approximate eigenstate of $\ket{\phi_q}$
for some $q$ such that
\be
 |c_q|^2 \equiv | \braket{\phi_q}{\psi}|^2 &=& p. 
\ee
Remembering that $\omega_q$ will be some real number between $0$ and $M$, we
define $\floor{\omega_q}$ to be the nearest $m$-bit integer less than
$\omega_q$, and $\ceil{\omega_q}$ to be the nearest $m$-bit integer greater
than $\omega_q$, where modulo $M$ has been assumed. The probability of
measuring the index register in either the
state $\ket{\floor{\omega_q}}$ or $\ket{\ceil{\omega_q}}$ is
\be
 P(\floor{\omega_q} \mathrm{or} \ceil{\omega_q})
 &=& \sum_k |c_k f(\omega_k,\floor{\omega_q}) |^2 \nonumber \\
 && \quad + \quad \sum_k |c_k f(\omega_k,\ceil{\omega_q}) |^2 \nonumber \\
 &\geq& |c_q f(\omega_q, \floor{\omega_q})|^2 +
    |c_q f(\omega_q, \ceil{\omega_q})|^2 \nonumber\\
 &\geq& |c_q|^2 \ 2\ \mathrm{sinc}^2(0.5) \nonumber\\
 &>& 0.8 p.
\ee
Hence, with probability greater than $0.8p$ we will obtain an approximate
eigenvalue associated with $\ket{\phi_q}$ which differs in phase from the actual
eigenvalue by less than $2 \pi/2^m$. 
Thus, if $p$ is reasonably large, we have a high probability of finding the
best estimate of the eigenvalue. However, as we shall see, large $p$ does not
imply that we will improve on the approximate eigenstate.

\begin{figure}[ht]
 \centering
 \scalebox{0.45}{\includegraphics{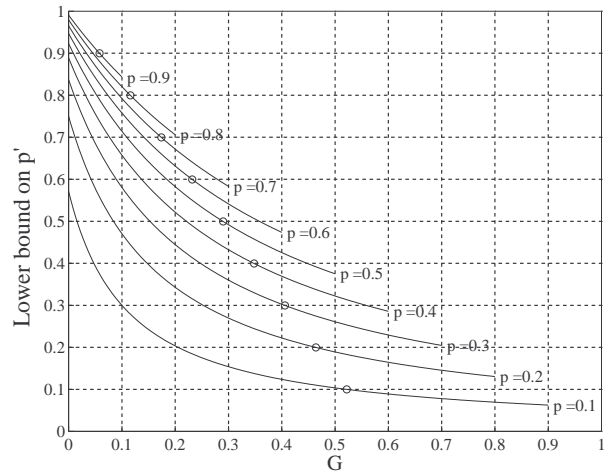}}
 \caption{The lower bound on $p'$ as a function of $G$ for
         $|f(\omega_q,[\omega_q])|^2 = 0.6$ and various values of $p$. The
     circles indicate the points at which the minimum of $p'$ equals $p$.}
 \label{minamp}
\end{figure}

Suppose we measure the index register in
the state $\ket{[\omega_q]}$, where $[\omega_q]$ denotes the closest $m$-bit
integer to $\omega_q$. (N.B. This will occur with probability greater than
$0.4p$, as $|f(\omega_q,[\omega_q])|^2 > 0.4$.) 
The key question that we wish to address in this paper is: has our initial
approximate eigenstate improved? 
Letting $p' \equiv |c'_q|^2$, we are effectively asking what bounds can be 
placed on $p'$? 
For an arbitrary $U$ it is obvious that the upper bound of $p' = 1$ can be 
obtained by setting $\ket{\psi} = \ket{\phi_q}$. 
We now investigate the lower bound by dividing the eigenstates into three
disjoint sets, 
\be\label{omegasets}
  \mathcal{Q} &=& \{q\}, \nonumber \\ 
  \mathcal{G} &=& \{g:g\not= q,|\omega_g-[\omega_q]| \leq 1\} \ \mathrm{and} \\
  \mathcal{H} &=& \{ h : |\omega_h - [\omega_q]| > 1 \}. \nonumber
\ee
We now have
\be
 p' &=& \frac{p |f(\omega_q,[\omega_q])|^2} {\mathcal{N}},
\ee
with
\be
 \mathcal{N} &=& p |f(\omega_q,[\omega_q])|^2 + 
    \sum_{g \in \mathcal{G}} |c_g|^2 |f(\omega_g,[\omega_q])|^2  \nonumber \\
   &&  \quad + \sum_{h \in \mathcal{H}} |c_h|^2 |f(\omega_h,[\omega_q])|^2. 
\ee
Using \eqns{phasefun}{omegasets}, it is not hard to show
\be\label{setlim}
  0.4 <  & \ \ |f(\omega_q,[\omega_q])|^2 \ \  & \leq 1 \nonumber \\ 
  0 \leq &|f(\omega_g,[\omega_q])|^2& \leq 1  \\ 
  0 \leq &|f(\omega_h,[\omega_q])|^2& \leq \lambda \nonumber  
\ee
where $\lambda = |f(1.5,0)|^2$. As $m$ increases $\lambda$ tends to
$(\frac{2}{3\pi})^2 \approx 0.045$. However for our analysis it is sufficient
to note that $0.045 < \lambda < 0.05$ for $m > 3$. \eqn{setlim} leads to the 
lower bound 
\be
 p' &\geq& \frac{p|f(\omega_q,[\omega_q])|^2}
   {p|f(\omega_q,[\omega_q])|^2 + (1-\lambda)G + \lambda(1-p)}
\ee
where $G = \sum_g |c_g|^2$. \fig{minamp} contains
a plot of this lower bound  as a function of $G$ for
$|f(\omega_q,[\omega_q])|^2 = 0.6$ and various values of $p$. The circles
indicate the points at which the minimum of $p'$ equals $p$.
Thus, we see that by endeavoring to make $G$ as small as possible, we increase
the amplitude of $\ket{\phi_q}$. 
For a given $U$ and $\ket{\psi}$, $G$ can be made arbitrarily small by
increasing $m$. However, we are interested in the performance of the algorithm
for small values of $m$.
We shall now look at $G$'s dependence on $U$ and $\ket{\psi}$ by attempting to
create eigenvectors for both the number and displacement operator in a ion trap.

\section{An Iontrap Implementation}\label{ionimp}

We first derive the Hamiltonian for $\Lambda(U)$, where $U$ is the evolution
operator associated with the number operator, and investigate the phase
estimation algorithm's performance for various initial states. We then derive
the Hamiltonian for the more complicated case of $U$ being the displacement
operator. For both of these examples the index register will be two
electronic levels of $m$ ions in a linear ion trap, and the target system will 
be the center-of-mass (CM) vibrational mode of the ions.

\subsection{The Number Operator}\label{numop}

Consider the standard Hamiltonian
of the one dimensional harmonic oscillator,
\be
 H = \hbar \omega (\adag a + \frac{1}{2}).
\ee
where $\adag$ and $a$ are the creation and annihilation operators.
Ignoring the over-all phase contribution of the zero energy state,
the unitary operator we will first be analyzing is
\be
 U(t) = e^{-i \omega \adag a t}.
\ee
In this case, $\Lambda(U)$ is given by 
\be\label{numLambda}
 \Lambda(U) &=& \exp\left({\frac{-it}{\hbar}\sum_{j=0}^{m-1}H_j}\right),
\ee
where
\be\label{Hnum}
 H_j &=& \hbar \adag a 2^j \omega (\sigma_z^{(j)} + \frac{1}{2}) .
\ee
The inversion operator for each ion is defined by $\sigma_z^{(j)} =
(\ket{0}\bra{0} - \ket{1}\bra{1})/2$.
This Hamiltonian can be obtained for interaction times greater than
the period of the CM vibrational mode by applying a set of
far-detuned standing wave pulses to the ion~\cite{D'Helon96}.

We begin our analysis by initializing the CM mode in some phonon number state
$\ket{n}$~\cite{Meekhof96}, and setting $\omega t = 2\pi(1 - 1/M)$.
It is important to note that we are assuming that all the higher vibrational
modes are in the vacuum state.
Assuming no errors, applying the phase estimation algorithm results in the
index register being measured in the state $\ket{n \mod M}$ and the target
system is left unchanged.
If we now let $\omega t$ be some arbitrary value, applying the algorithm will
leave the target system unchanged, and the index system will be measured in
the state $\ket{j}$ with probability
\be
 P(j) &=& \left|f\left( \frac{-\omega t n M}{2\pi}, j\right)\right|^2
\ee

Let us consider the more interesting situation where the target system is
initialized in some coherent state $\ket{\alpha}$. We can utilize the phase
estimation algorithm to transform the state of the target system into an
approximation to a Fock state. 

\begin{figure}[ht]
 \centering
 \scalebox{0.45}{\includegraphics{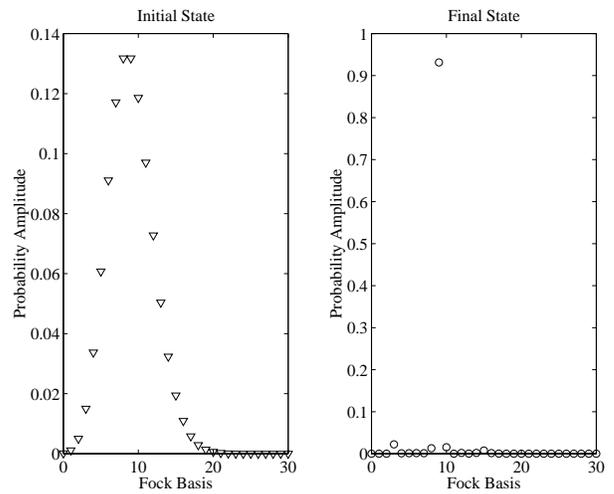}}
 \caption{Fock state distributions for the target system initially in a
 coherent state with $\alpha = 3$, and the state of the system after applying
 the phase estimation algorithm and measuring the four index qubits.}
 \label{fock}
\end{figure}

For example, suppose we use four index qubits,  $\omega t = 1$ and we choose 
to approximate the Fock state $\ket{n = 9}$ by using the coherent 
state $\ket{\alpha = 3}$. 
In this example we perhaps might think that $\ket{\alpha = 3}$ is not a good 
approximate state because $p \approx 0.13$ however the fact that $G < 0.035$
indicates the algorithm should work well. Applying the algorithm and measuring 
the index register in the state $\ket{9}$, we obtain $p' \approx 0.93$.
The initial and final target state for this scenario is shown in \fig{fock}.

Having shown that the phase estimation algorithm can be used to generate Fock
states from coherent states, we now attempt to generate eigenstates of the  
displacement operator.

\subsection{The Displacement Operator}

The displacement operator applied to the CM vibrational mode is defined as
\be\label{disop}
 D(\alpha) &=& \exp(\alpha \adag - \alpha^* a).
\ee
Thus the operator we wish to apply is
\be\label{disLambda}
 \Lambda(D) &=& \exp\left({\frac{-it}{\hbar}\sum_{j=0}^{m-1}H_j}\right),
\ee
where the $H_j$ are now defined as
\be\label{disHam}
 H_j &=& i \hbar (\alpha\adag - \alpha^* a) 2^j (\sigma_z^{(j)} + \frac{1}{2}).
\ee
It has already been shown~\cite{Monroe96}
that conditional displacement operations such as the Hamiltonian in
\eqn{disHam} can be performed in an ion trap.

It is not hard to show that 
\be
  D(de^{i(\phi+\frac{\pi}{2})}) \ket{\alpha, \varepsilon}
  &\approx&  e^{i2d|\alpha| e^{-r}} \ket{\alpha, \varepsilon}
\ee
for large values of squeezing parameter $r$ and where $\alpha = |\alpha|
e^{i\phi}$, $\varepsilon = r e^{2i\phi}$ and
\be
  \ket{\alpha, \varepsilon} &\equiv& 
  S(\varepsilon) D(\alpha) \ket{0}
\ee
is a squeezed coherent state. Thus the squeezed coherent states 
$\ket{\alpha, \varepsilon}$ form approximate eigenvectors of the 
displacement operator $D(de^{i(\phi+\frac{\pi}{2})})$. 

Without loss of generality we can set $\phi = 0$ in which case the eigenstates
of the displacement operator are simply the position eigenstates. It is then
not hard to show that for small fixed $m$, $G \approx 1-p$, which leads 
to $p' \approx p$. Thus applying the phase estimation algorithm to squeezed
displaced states does not produce improved eigenstates of the displacement
operator. 

\section{Conclusion}\label{conclusion}

We have shown that the phase estimation algorithm can be used to generate
eigenstates of the number operator, even when we severely limit the size 
of the index system.
It would be interesting to see if an analogous implementation could be
performed using cavity QED, allowing generation of photon number states with
only small numbers of trapped atoms.
We have also shown that the algorithm's performance depends on the relation 
between the approximate eigenstate and the spectrum of the operator. 
We can gauge the algorithm's performance by calculating a parameter $G$. 

\acknowledgments
B.~C.~T. acknowledges support from the University of Queensland Traveling
Scholarship, and thanks to S.~Lloyd, S.~Schneider, M.~Nielson and 
D.~F.~V.~James for helpful discussions.



\end{document}